\global\def\draftcontrol{0}

   \def\versionno{Disordered Universe, Intro -- draft   }

\catcode`\@=11

\expandafter\ifx\csname draftcontrol\endcsname\relax\global\def\draftcontrol{0}
\fi

{\count255=\time\divide\count255 by 60
\xdef\hourmin{\number\count255}
\multiply\count255 by-60\advance\count255 by\time
\xdef\hourmin{\hourmin:\ifnum\count255<10 0\fi\the\count255}}
\def\draftdate{\number\month/\number\day/\number\year\ \ \ \hourmin }

\newcommand\makepapertitle{\par
  \begingroup
    \renewcommand\thefootnote{\@fnsymbol\c@footnote}%
    \def\@makefnmark{\rlap{\@textsuperscript{\normalfont\@thefnmark}}}%
    \long\def\@makefntext##1{\parindent 1em\noindent
            \hb@xt@1.8em{%
                \hss\@textsuperscript{\normalfont\@thefnmark}}##1}%
     \newpage
     \global\@topnum\z@   
     \@makepapertitle
     \thispagestyle{empty}\@thanks
  \endgroup
  \setcounter{footnote}{0}%
  \global\let\thanks\relax
  \global\let\makepapertitle\relax
  \global\let\@makepapertitle\relax
  \global\let\@thanks\@empty
  \global\let\@author\@empty
  \global\let\@date\@empty
  \global\let\@title\@empty
  \global\let\title\relax
  \global\let\author\relax
  \global\let\date\relax
  \global\let\and\relax
  \def\version{\let\version\@version\@gobble}
}
\def\@makepapertitle{%
  \newpage
   \ifnum\draftcontrol=1 {}
   \version\versionno
   \vskip 3em%
   \else
   \hfill\hbox to 3cm {\parbox{4cm}{\@pubnum}\hss}%
   \vskip 3em%
   \fi
   \begin{center}%
   \let \footnote \thanks
     {\LARGE {\@title}}%
     \vskip 1.5em%
     {\normalsize
       \lineskip .5em%
       \begin{tabular}[t]{l}%
         \@author
       \end{tabular}\par}%
     \vskip 1.5em%
     {\@bstract}%
     \end{center}%
     \vskip 1.5em
     \@date%
   \par
}

\gdef\@pubnum{}
\def\pubnum#1{%
  \gdef\@pubnum{#1}}

\gdef\@bstract{}
\def\Abstract#1{%
  \gdef\@bstract{%
   \parbox{\textwidth-0pc}{%
   \centerline{\bf Abstract}\penalty1000%
\kern.2cm%
\noindent
\renewcommand\baselinestretch{1.0}%
{#1}}}
}

\def\ps@paper{\let\@mkboth\@gobbletwo%
     \ifnum\draftcontrol=1
    \def\@oddfoot{\hbox to \textwidth{\tiny \versionno \hfil\tiny\draftdate}%
    \hskip -\textwidth \hbox to \textwidth{\hfil\rm\thepage\hfil}}%
     \else\def\@oddfoot{\hbox to \textwidth{\hfil\rm\thepage\hfil}}
     \fi
     \let\@evenfoot\@oddfoot
}

\def\body{\clearpage
          \pagestyle{paper}
    }
\newenvironment{acknowledgements}{%
\vskip 3.25ex
\noindent {\bf Acknowledgements}
}

\def\@version#1{\ifnum\draftcontrol=1
\typeout{}\typeout{#1}\typeout{}
\vskip3mm\centerline{\hbox{\fbox{\normalsize{\tt DRAFT -- #1 -- }
                   {\draftdate}}}}\vskip3mm
\fi}
\let\version\@version
\long\def\eqlabel#1{\ifnum\draftcontrol=1
                    \tag@false  
                    \tag*{(\theequation) \hbox to -0.2cm{\hspace{0cm}\small{#1}\hss}}
                    \refstepcounter{equation}
                    \edef\@currentlabel{\theequation}
                    \ltx@label{#1}          
                    \else
                    \label{#1}
                    \fi
                    }
\let\st@bibitem\@bibitem
\let\st@lbibitem\@lbibitem
\ifnum\draftcontrol=1
  \def\@bibitem#1{%
    \st@bibitem{#1}\a@@label{#1}\ignorespaces}
  \def\@lbibitem[#1]#2{%
    \st@lbibitem[#1]{#2}\a@@label{#2}\ignorespaces}
  \def\a@@label#1{%
    \gdef\a@lab{\smash{\normalfont\small#1}}
    \ifvmode
      \if@inlabel
        \global\setbox\@labels\hbox{%
          \llap{\a@lab\let\a@lab\relax
                \kern\@totalleftmargin\kern\marginparsep}%
          \box\@labels}%
      \fi
    \fi}
\fi

\documentclass[10pt]{article}

\usepackage{amsmath,amssymb,array,calc,epsfig,bm}
\usepackage{graphicx}        
\usepackage{multicol}        

\usepackage{amsmath,amssymb,bm,epsfig}

\usepackage{cite}
\usepackage[
      colorlinks=true,
      linkcolor=red,  
      urlcolor=blue,    
      filecolor=red,     
      linkcolor=blue,
      citecolor = red,
      pdfstartview=FitV,
      pdftitle={},
        pdfauthor={Paolo Benincasa},
        pdfsubject={},
        pdfkeywords={},
        pdfpagemode=None,
        bookmarksopen=true    
      ]{hyperref}

\usepackage{centernot}      
\usepackage{mathpazo}
\usepackage{stmaryrd}
\usepackage{pifont}
\usepackage{colortbl}
\usepackage{multirow,bigdelim}
\usepackage{arydshln}
\usepackage{tikz, pict2e}
\usepackage{geometry}
\usepackage{lmodern}
\usepackage{wrapfig}
\usepackage{float}
\usepackage{enumitem}
\usepackage[framemethod=default]{mdframed}
\usepackage{array}
\usepackage{mathrsfs}

\usepackage{placeins}
\usepackage{circuitikz}

\usepackage[T1]{fontenc}
\usepackage[utf8]{inputenc}
\usepackage[english]{babel}

\PassOptionsToPackage{table}{xcolor}

\usepackage{xspace}
\usepackage{ifthen}
\usepackage{epstopdf}
\usepackage{amsthm}

\usepackage{relsize}
\usepackage{tabularx}
\usepackage{booktabs}
\usepackage{soul}
\usepackage{pdfpages}
\usepackage[most]{tcolorbox}
\usepackage{capt-of}
\usepackage{xfrac}
\usepackage{empheq}
\usepackage{slashed}
\usepackage{cancel}
\usepackage{upgreek}
\usepackage[skip=1em]{parskip}

\usepackage{latexsym}
\usepackage{ifoddpage}

\usepackage{subcaption}

\usepackage{cancel}
\usepackage{comment}
\usepackage{shuffle}

\usepackage{savesym}
\savesymbol{Ref}
\savesymbol{erf}
\usepackage{physics}
\restoresymbol{phys}{Ref}
\restoresymbol{phys}{erf}

\usepackage[compat=1.1.0]{tikz-feynman}
\tikzfeynmanset{compat=1.1.0}

\definecolor{myblue}{RGB}{166, 193, 222}
\definecolor{myblue2}{RGB}{68, 114, 196}
\definecolor{mygreen}{RGB}{175, 218, 162}
\definecolor{mygreen2}{RGB}{133, 210, 110}
\definecolor{myorange2}{RGB}{255, 192, 0}

\newtcolorbox{shade}[1][]{
  top=-7pt,bottom=6pt,
  sharp corners=all,
  colback=white!95!black,
  colframe=white,
}

\newcolumntype{P}[1]{>{\centering\arraybackslash}p{#1}}
\newcolumntype{L}{>{$}l<{$}}

\usetikzlibrary{calc,backgrounds}
\usetikzlibrary{arrows,shapes,positioning,shadows,trees}
\usetikzlibrary{arrows.meta, shapes.geometric}
\usetikzlibrary{mindmap}
\usetikzlibrary{decorations.pathreplacing}
\usetikzlibrary{decorations.pathmorphing}
\usetikzlibrary{decorations.markings}
\usetikzlibrary{matrix}
\usetikzlibrary{patterns.meta}

\tikzstyle arrowstyle=[scale=1]
\tikzstyle directed=[postaction={decorate,decoration={markings,
    mark=at position .5 with {\arrow[arrowstyle]{stealth}}}}]
\tikzstyle reverse directed=[postaction={decorate,decoration={markings,
    mark=at position .5 with {\arrowreversed[arrowstyle]{stealth};}}}]

\ifnum\draftcontrol=1
\fi

\renewcommand\baselinestretch{1.25}
\renewcommand\section{\@startsection {section}{1}{\z@}%
                                   {-3.5ex \@plus -1ex \@minus -.2ex}%
                                   {2.3ex \@plus.2ex}%
                                   {\normalfont\large\bfseries}}
\renewcommand\subsection{\@startsection{subsection}{2}{\z@}%
                                   {-3.25ex\@plus -1ex \@minus -.2ex}%
                                   {1.5ex \@plus .2ex}%
                                   {\normalfont\normalsize\bfseries}}
\renewcommand\subsubsection{\@startsection{subsubsection}{3}{\z@}%
                                   {-3.25ex\@plus -1ex \@minus -.2ex}%
                                   {1.5ex \@plus .2ex}%
                                   {\normalfont\normalsize\it}}
\renewcommand\paragraph{\@startsection{paragraph}{4}{\z@}%
                                   {-3.25ex\@plus -1ex \@minus -.2ex}%
                                   {1.5ex \@plus .2ex}%
                                   {\normalfont\normalsize\bf}}

\numberwithin{equation}{section}

\def\revise#1       {\raisebox{-0em}{\rule{3pt}{1em}}%
                     \marginpar{\raisebox{.5em}{\vrule width3pt\
                     \vrule width0pt height 0pt depth0.5em
                     \hbox to 0cm{\hspace{0cm}{%
                     \parbox[t]{4em}{\raggedright\footnotesize{#1}}}\hss}}}}

\def\sqr#1#2{{\vcenter{\vbox{\hrule height.#2pt
 \hbox{\vrule width.#2pt height#1pt \kern#1pt
 \vrule width.#2pt}\hrule height.#2pt}}}}

\catcode`\@=12

\newcommand{\ee}{e}

\newcommand{\beq}{\begin{equation}}
\newcommand{\eeq}{\end{equation}}
\newcommand{\bea}{\begin{equation}\begin{aligned}}
\newcommand{\eea}{\end{aligned}\end{equation}}

\newlength{\wsingfig}
\newlength{\wdblefig}
\newlength{\wquadfig}
\newlength{\wtriplefig}
\newcommand   \lptl{\raise .8ex\hbox{$^\leftarrow$} \hspace{-9pt} \partial}
\newcommand   \lrptl{\raise .8ex\hbox{$^\leftrightarrow$} \hspace{-9pt} \partial}

\newcommand   \be {\mathbf{e}}

\def\be#1\ee{\begin{align}#1\end{align}}

\newcounter{ssooeexercise}[section]
\renewcommand{\thessooeexercise}{\thesection.\arabic{ssooeexercise}}

\newtcolorbox{exbox}{
  colback=gray!10,
  colframe=gray!20,
  boxrule=1pt,
  arc=5pt,
  breakable,
  left=1em,
  right=1em,
  top=0.8em,
  bottom=0.8em
}

\newcommand{\newexercise}[3][]{%
  \refstepcounter{ssooeexercise}%
  \expandafter\gdef\csname ex@content@#2\endcsname{#3}%
  \expandafter\xdef\csname ex@number@#2\endcsname{\thessooeexercise}%
  \begin{exbox}%
    \textbf{Exercise~\thessooeexercise:} #3
    \ifx&#1&\else\label{#1}\fi
  \end{exbox}%
}

\newcommand{\printexercise}[1]{%
  \begin{exbox}%
    \textbf{Exercise~\csname ex@number@#1\endcsname:}
    \csname ex@content@#1\endcsname
  \end{exbox}%
}

\begin{document}


\title{\underline{\Large\bf The Disordered Universe 2025:} \\
       {\Large\bf An intro and a vision}}

\pubnum{%
}

\author{%
	{\bf Paolo Benincasa}${}^{\dagger\,\ddagger}$, {\bf Guilherme Leite Pimentel}${}^{\diamond}$
    	\\ [0.4cm]
    	\ttfamily ${}^{\dagger}$ Asociaci{\'o}n Cultural La Ricotta, Espazo La Ricotta\\
    	\ttfamily $\phantom{{}^{\dagger}}$ As Barreiras, Castro Caldelas, E-+32769 Galicia-Spain\\
    	\ttfamily ${}^{\ddagger}$ Instituto Galego  de F{\'i}sica de Altas Enerx{\'i}as IGFAE, \\
    	\ttfamily $\phantom{{}^{\dagger}}$ Universidade de Santiago de Compostela, E-15782 Galicia-Spain\\
    	\ttfamily ${}^{\diamond}$ Scuola Normale Superiore, \\
    	\ttfamily $\phantom{{}^{\dagger}}$ Piazza dei Cavalieri 7, 56126, Pisa, Italy\\[0.4cm]
    	\small\tt pablowellinhouse@sindominio.net, guilherme.leitepimentel@sns.it \\ [0.75cm]
}

\Abstract{%
The {\it Disordered Universe} is a novel research direction that exploits ideas and techniques coming from apparently unrelated research areas, ranging from cosmology to scattering amplitudes in particle physics and disordered complex systems, with the aim of addressing open questions in cosmology at all scales, from the very infra-red (or very late time) to the early universe. This paper acts as a general introduction and vision for a set of lectures delivered at the first \href{https://indico.sns.it/event/91/}{\it La Ricotta Summer School ``The Disordered Universe 2025''}, held from 21 July to 10 August 2025. These lectures involved open quantum systems, analytic properties of scattering amplitudes and correlation functions, foundations and quantum aspects of the de Sitter universe, as well as statistical physics out of equilibrium and spin glasses. While the lecture notes related to each course taught at the school will be published separately and are inevitably devoted to both conceptual and technical aspects of each of these topics, the school was not meant to be a {\it collage} of independent topics: the present paper has the purpose of outlining the global vision, how cosmology ends up being a unifier and benefits from both their viewpoint and techniques, how all the concepts and techniques originally coming from these four different areas are part of the same cosmological narrative, why this point of view deserves to be explored, and drawing perspectives for the future. We consider it crucial to build a microscopic theory for the early universe and maybe one that embraces all scales. 
}

\makepapertitle

\body

\version\versionno

\newpage

\section*{Preface}
This article combines the introduction and conclusion chapters of the book {\it The Disordered Universe: Open Systems, Scattering Amplitudes, de Sitter Space, and Disordered Systems in Cosmology}~\cite{Benincasa:2027tdu}, which collects the lectures delivered at the \href{https://indico.sns.it/event/91/}{\it La Ricotta Summer School: The Disordered Universe 2025}, held from 21~July to 1~August~2025 at the \href{https://blogs.sindominio.net/laricotta/#}{\it Espazo La Ricotta} in As Barreiras (Castro Caldelas, Spain). These lecture notes are intended as a pedagogical introduction to topics that are becoming increasingly relevant in cosmology. 

\newpage

\tableofcontents


\section{Background and general motivation} 

Cosmology poses some of the deepest questions in science, from the origin and evolution of all the structures we can observe in the sky to the origin of fundamental notions like space and time which appear to be ingrained in the human mind. Our current understanding seems to hint at a fascinating connection between the large and small scale physics, mechanisms characterising statistical systems as well as the need for a novel mathematical language, to be able to address them.

All the structure in our universe has evolved from a hot and dense state, named the {\it Hot Big Bang}, and, hence, anything we can observe acts as a window on the past of our universe, and must contain a signature of the physics at that stage of the universe: for example, the distribution of galaxies in the sky can be traced back to patterns in the density variations at the origin of the Hot Big Bang, which are referred to as {\it primordial fluctuations}. 

\begin{figure}[t]
	\centering
	\includegraphics[width=.675\textwidth, height=.675\textheight, keepaspectratio]{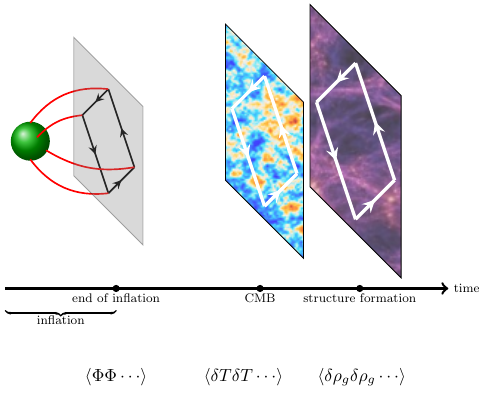}
	\caption{{\it Cartoon of the evolution of the Universe --}  It highlights three of its phases: the end of inflation, where correlation functions $\langle\Phi\Phi\cdots\Phi\rangle$ of quantum states encode the physical processes occurred during inflation; the CMB becoming visible, from which we can observe patterns, such as the correlations $\langle\delta T\delta T\cdots\delta T\rangle$ for temperature fluctuations; the large scale structure formation, which allows us to extract information about our universe from their density distributions $\langle\delta\rho_g\delta\rho_g\cdots\delta\rho_g\rangle$. The correlations $\langle\delta T\delta T\cdots\delta T\rangle$ and $\langle\delta\rho_g\delta\rho_g\cdots\delta\rho_g\rangle$ can be traced back to the end of the inflation where $\langle\Phi\Phi\cdots\Phi\rangle$ is the initial condition for their evolution and constitutes a portal to understand what we observe and to extrapolate the early universe physics.}
\end{figure}

Recent cosmological observations have revealed the remarkable fact that these primordial fluctuations must have been created before the hot Big Bang: regions of space which are not causally connected have been shown to be impressively homogeneous. This {\it horizon problem} can be solved by hypothesising that the conventional Big Bang was therefore not the beginning of time, rather the end of an earlier high-energy period, and the primordial fluctuations were created in this pre-Big Bang phase -- see \cite{Baumann:2009ds} and references therein.  The leading proposal for the resolution of this problem, named {\it inflation}, postulates a phase of very rapid, accelerated, expansion in the very early universe. During this inflationary period, small quantum fluctuations were stretched to cosmic scales and became the density fluctuations which seeded the formation of stars and galaxies -- see \cite{Baumann:2009ds} and references therein. Although this provides an elegant explanation for the primordial seed fluctuations, the physics of inflation remains one of the biggest open problems in fundamental physics, as it lacks a complete framework. The upper bound on the energy during inflation is estimated to be $10^{14}$ GeV \footnote{This estimation can be obtained by thinking that the tensor-scalar ratio $r:=A_t/A_s$ between the amplitudes of the tensor $A_t$ and scalar $A_s$ perturbations, can be written in terms of the Hubble parameter $H$ and the Planck mass $M_{\mbox{\tiny\scshape pl}}$ as
\begin{equation}
	r\:\sim\:\frac{16}{A_s\pi}\left(\frac{H}{M_{\mbox{\tiny\scshape pl}}}\right)^2.
\end{equation}
From this expression, $H$ acquires the form
\begin{equation}
	H\:\sim\:r\frac{M_{\mbox{\tiny\scshape pl}}\sqrt{\pi A_s}}{4}
\end{equation}
and, hence, considering that the parameters $r$ and $A_s$ have been bounded by observations to be $r\:\lesssim\:0.032-0.037$ \cite{Tristram:2021tvh, BICEP:2021xfz} and $A_s\:=\:(2.10\pm 0.03)\cdot 10^{-9}$ \cite{Planck:2018vyg}, one obtains the upper bound of $10^{14}$ GeV.}, ten orders of magnitude larger than the one reached in the biggest experiment on Earth, the Large Hadron Collider. Furthermore, physics at such an enormous energy brings us tantalisingly close to the Planck scale -- {\it just} five orders of magnitude less --, where our knowledge completely breaks down and where the true answer about the origin of our Universe and the basic principles ruling the physics in it are encoded.

An intriguing feature of inflation is that our view of this epoch is frozen in time. We can only make inferences about the inflationary era from spatial correlations of the primordial fluctuations in the post-inflationary universe. These primordial correlations live on the future boundary of the inflationary space-time or, equivalently, the past boundary of the hot Big Bang universe. This also means that they encode the low energy imprint of the high energy inflationary processes. Learning how to extract the dynamics of inflation from these static boundary correlations becomes then an essential goal: it would allow us to understand the imprint of inflation in the structures we can observe, and have a handle on {\it ultra high energy} processes and, thus, a concrete bridge to the mysterious Planck scale.

The standard approach to compute inflationary correlators is to follow their time evolution from early times to the end of inflation. However, this requires assumptions about the interactions during inflation, and often leads to overly complicated computations. Alternatively, we can try to understand which principles constrain the inflationary physics and, thus, which rules the inflationary correlation functions ought to satisfy. Reaching such an understanding would be of paramount importance as these constraints can function both as model-independent predictions for the observations of early-time physics, and as consistency conditions on the inflationary proposal itself.

\begin{equation*}
    \begin{tikzpicture}
            \node[draw, rectangle, rounded corners, thick, color=blue!80, fill=blue!10, text width=14.875cm, text height=2cm] (O1) at (0,0) {};
            \node[font=\it, align=justify, text width=14.75cm] (O1txt) at (O1) {%
            	\color{black}{The basic premise behind the {\scshape disordered universe} is that a framework for inflation will necessarily} {incorporate developments from the field of out-of-equilibrium}  {complex systems, as well as modern methods in particle physics}%
            };
    \end{tikzpicture}
\end{equation*}

Recent efforts -- see \cite{Benincasa:2022gtd} for a review and references for the original literature -- have shown how ideas and tools from particle physics and disordered systems alike can provide important insights: Modern methods for flat-space scattering amplitudes, together with the identification of an underlying combinatorial structure for perturbative processes \cite{Arkani-Hamed:2017fdk, Arkani-Hamed:2018bjr, Benincasa:2018ssx, Benincasa:2019vqr, Benincasa:2020uph, Benincasa:2021qcb, Albayrak:2023hie, Benincasa:2024leu, Benincasa:2024lxe, Benincasa:2024rfw, Arundine:2026fbr, Arundine:2026myr}, have shed light on the computation and physical content of inflationary correlation functions, while the application of overlap distributions stemming from the theory of spin glasses \cite{Parisi:1983dx, Mezard:1986mpv, Parisi:2019puz} has led to an ultrametric organisation of the late-time configurations space of light quantum states in an inflationary background \cite{Anninos:2011kh, Roberts:2012jw, Shaghoulian:2013qia, Benincasa:2026bcl}. The consistency conditions and principles uncovered by studying primordial fluctuations, along with the necessity to incorporate gravitational effects at quantum level during inflation, may deepen our understanding of what constitutes a truly consistent inflationary theory. When combined with the conjectured equivalence, known as {\it holography}, between non-gravitating quantum systems and theories of quantum gravity -- see \cite{Aharony:1999ti} and references therein --, insights suggest that space-time, from a microscopic quantum perspective, emerges collectively from a large number of degrees of freedom.

\begin{figure}[t]
	\centering
	\includegraphics[width=.375\textwidth, height=.375\textheight, keepaspectratio]{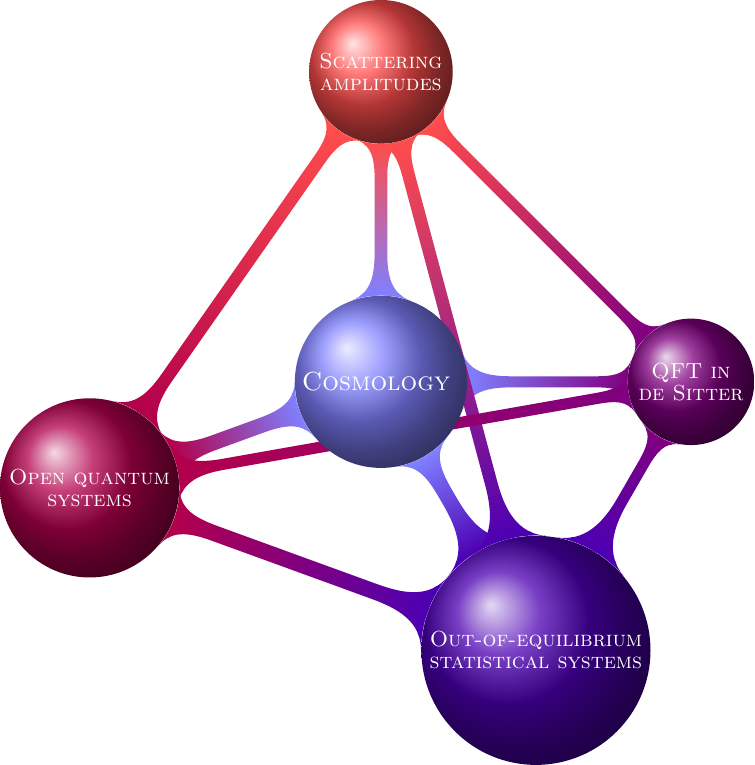}
	\caption{{\it Interplay for cosmology --}  Unveiling the early universe physics requires being able to extract information out of physical observables. Thinking alternatively of the universe and cosmology as a closed or open system allows a wide spectrum of handles on this issue with the understanding of the analytic structure of the relevant observables, following the scattering amplitude lesson as a way to extract information out of them and tying it to physical principles. However, what should we compute? What are the quantities which precisely contain such information? A careful understanding of how to pack the degrees of freedom in an expanding universe, the peculiar feature of a $\Lambda\,>\,0$ universe, and the understanding of the state space offers a new way  to identify new observables, following the spin-glass lessons.}
\end{figure}

While these developments are indeed exciting, they have clearly just scratched the surface of an important topic, and we have only started to develop the tools to answer deep physical questions. It thus becomes a general aim to develop a consistent framework for addressing such important issues, and the \href{https://indico.sns.it/event/91/}{La Ricotta Summer School {\it The Disordered Universe 2025}} and its lecture notes associated to this paper -- everything recollected in a book in press \cite{Benincasa:2027tdu} --, aims to set the stage for this to happen in the not-too-distant future, maybe with the important contribution by one or more among the participants themselves of the school, or by anyone else who decides to go through these lecture notes.

\section{Understanding cosmological observables: a pathway} 

It is important to remark that the languages we are accustomed to talk about physical phenomena at {\it accessible high energy} scales -- {\it i.e.} the ones probed so far by our experiments -- are Quantum Field Theory (QFT), which makes the principles of quantum mechanics and special relativity manifest \cite{Weinberg:1995mt}, and General Relativity (GR), whose Riemannian geometry formulation provides a realisation of the equivalence principle \cite{Weinberg:1972kfs}. They provide a narrative, through unphysical degrees of freedom, which connects initial and final (physical) states in a physical process. The great success of these frameworks came with a price. 

Many of the phenomena they describe appeared, until recent years, as some sort of magic arising from complicated calculations. The prototypical example is given by the Parke-Taylor formula for tree-level scattering of gluons \cite{Parke:1986gb}, whose simplicity is completely obscured by the usual Feynman diagram computation. Making the first principles behind particle physics manifest, as Feynman diagrams and more generally the field theory description do, has resulted in very cumbersome computations that obscured both much of the structure of the theories and the origin of the phenomena they predict. Improving our knowledge of the analytic structure of physical observables and how it is tied to the fundamental principles behind flat-space physics has been of crucial importance to achieving a deeper understanding of particle physics, developing higher predictive power and more powerful computational techniques. And this is the first lesson that can be carried through to cosmology from the study of {\it scattering amplitudes} in particle physics. What are these first principles that rule particle interactions? 
\begin{itemize}
	\item[\ding{111}] The {\it unitarity} of time evolution: in loose words, it is the statement that probabilities have to sum up to one and they have to be individually positive; 
	\item[\ding{111}] The {\it cluster decomposition principle}: it is the idea that when two events are taken away from each other, their mutual influence by exchanging information decreases sufficiently fast with the distance;
	\item[\ding{111}] {\it A relativity principle}, which in particle physics is provided by the invariance under the Lorentz group. 
\end{itemize}
They turn out to constrain the mathematical structure of scattering processes, providing in turns constraints on the processes which can actually occur. For example, all possible three particle couplings \cite{Benincasa:2007xk, Arkani-Hamed:2017jhn}, the consistency of interacting theories with a finite number of massless particles with spin less or equal to $2$ \cite{Weinberg:1964ew, Benincasa:2007xk, Benincasa:2011kn, Benincasa:2011pg, Arkani-Hamed:2017jhn}, charge conservation and the equivalence principle \cite{Weinberg:1964ew}, the graviton uniqueness theorem \cite{Benincasa:2007xk}, the Yang and Weinberg-Witten theorems \cite{Arkani-Hamed:2017jhn} are all instances of the simultaneously validity of these principles. And this goes further. In order to fix the ideas, let us define an S-matrix operator $\hat{S}$ which is unitary and Lorentz invariant by construction
\begin{equation}
	\hat{S}\: :=\:e^{i\lambda\hat{\mathcal{O}}},\qquad
	\mbox{with } \hat{\mathcal{O}}\;\Big|\langle 3,4|\hat{\mathcal{O}}|1,2\rangle\:=\:
	\delta^{\mbox{\tiny $(4)$}}\left(p_1+p_2-p_3-p_4\right)
\end{equation}
where $\lambda$ is just a coupling constant. We can then consider a $2-2$ scattering and expand in perturbation theory, obtaining
\begin{equation}\label{eq:22s}
	\langle 3,4|\hat{S}|1,2\rangle\:=\:\delta^{\mbox{\tiny $(4)$}}\left(p_1+p_2-p_3-p_4\right)
	\left\{
		1\:+\:i\lambda\:+\:\lambda^2\vartheta(s-4m^2)\:+\:\ldots
	\right\}
\end{equation}
where $\vartheta(s-4m^2)$ is the Heaviside step function and $s:=(p_1+p_2)^2$. Is there anything wrong with equation (\ref{eq:22s})? 

\begin{figure}[t]
	\centering
	\includegraphics[width=\textwidth, height=\textheight, keepaspectratio]{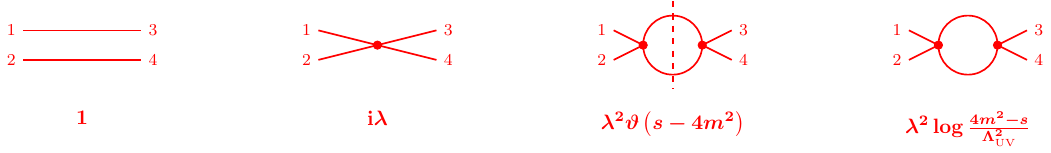}
	\caption{{\it Unitary and Lorentz invariant $S$-matrix and cluster decomposition violation --}  The first three graphs represent the $S$-matrix operator $\hat{S}:=e^{i\lambda\hat{\mathcal{O}}}$ is defined in (\ref{eq:22s}) in a perturbative expansion up to order $\lambda^2$. The first contribution is the free propagation, the second one the tree level/contact interaction, while the third one shows a distribution and it can be associated to a cut graph, constituting a violation of the cluster decomposition principle. The contribution compatible with cluster decomposition is a bubble graph, in the fourth term in the picture, whose associated function is a logarithm -- $\Lambda_{\mbox{\tiny UV}}$ is a UV cut-off due to the presence of UV divergences in such a graph.}
\end{figure}

While the first term in it represents the free propagation and the second one is just the contact interaction, both of which do not show any type of problem, the order $\lambda^2$ term is pathological as it turns out not to satisfy the cluster decomposition principle: in momentum space, such a principle translates into the statement that a scattering process can be represented by functions that, modulo the overall momentum conserving $\delta$-function, are analytic everywhere, with the only points of non-analyticity that can occur being poles and branch cuts. Hence, a distribution such as the Heaviside-$\vartheta$ is not allowed. In fact, the $\mathcal{O}(\lambda^2)$ term corresponds to a bubble diagram with all its propagators on-shell, rather than to an actual physical process, which would be expressed in terms of a logarithm. More explicitly, knowing that cluster decomposition does not allow for anything worse than poles and branch cuts would be enough to recognise (\ref{eq:22s}) as pathological, and this type of statement becomes completely {\it model independent}. There is one more lesson that formula \eqref{eq:22s} teaches us for cosmology: it is important to have control over higher order in perturbation theory, irrespectively of whether such correction are observable or not, as a consistency check on the tree level (or anyhow lower order) predictions and on the model itself. 

This goes even further as there exist some relations, connected to causality and unitarity, known as the Steinmann relations, which allow one to further constrain the analytic structure of scattering processes, even allowing them to be bootstrapped. Originally found as relations among correlators in different kinematic regions \cite{Steinmann:1960soa, Steinmann:1960sob, Araki:1961hb, Ruelle:1961rd}, they can be stated as the incompatibility of the double discontinuities along two partially overlapping channels in the physical region of scattering amplitudes \cite{Stapp:1971hh, Cahill:1973px, Cahill:1973qp, Lassalle:1974jm}, {\it i.e.} in this region the double discontinuities along this class of channels vanish, but it can be generalised to the study of the compatibility/incompatibility of all sequential discontinuities \cite{Bourjaily:2020wvq, Benincasa:2020aoj, Benincasa:2021qcb}. The knowledge of the compatible sequential discontinuities can be complemented, in the case of individual integrals, with the knowledge of the loci where their singularities can arise, which are determined by the Landau equations \cite{Nakanishi:1959jzx, Landau:1959fi, Bjorken:1959fd}, and the knowledge about the classes of functions which are expected: all together they identify a finite space of functions that the individual Feynman integrals are expected to evaluate to \cite{Hannesdottir:2024hke}.

Such a knowledge and point of view is not yet available in cosmology, but in the last nine years it started to be developed, from the idea of constraining early-universe processes using first principles \cite{Arkani-Hamed:2018kmz, Sleight:2019hfp, Baumann:2019oyu, Baumann:2020dch, Pajer:2020wnj, Pajer:2020wxk, Jazayeri:2021fvk, Hogervorst:2021uvp, DiPietro:2021sjt, Baumann:2022jpr, Benincasa:2022gtd, Wang:2022eop, DuasoPueyo:2023kyh}, to the investigation and exploitation of their analytic structure to compute them \cite{Goodhew:2020hob, Benincasa:2020aoj, Melville:2021lst, Baumann:2021fxj, Meltzer:2021zin, Cabass:2021fnw, Benincasa:2021qcb, Benincasa:2022gtd, Albayrak:2023hie}. However, this is just the beginning of the road. 

\begin{equation*}
    \begin{tikzpicture}
            \node[draw, rectangle, rounded corners, thick, color=blue!80, fill=blue!10, text width=14.875cm, text height=4cm] (O2) at (0,0) {};
            \node[font=\it, align=justify, text width=14.75cm] (O1txt) at (O2) {%
            \color{black}{Understanding the analytic structure of cosmological observables becomes of crucial importance both for extracting} {physics out of them via compatibility with first principles, and to}  {provide novel computational tools that make it possible to tame higher orders in perturbation theory.} {Although many of these corrections might not be observable, the lowest orders among them provide general consistency conditions, while} {the higher orders can provide a consistency check on the tree-level predictions.} 
            {The chapter {\scshape Analytic Properties of Amplitudes and Correlators} \cite{McLeod:2026tdu} offers a pedagogical discussion of the most recent} {developments in understanding the analytic structure of physical observables, aiming to boost progress in this direction.}%
            };
    \end{tikzpicture}
\end{equation*} 

\section{Getting disordered}\label{sec:Dis}

If at accessibly high energies cluster decomposition is one of the basic principle for constructing our theories and which constrains the analytic structure of scattering amplitudes, in expanding universes correlation functions of massless and sufficiently light states fail to cluster -- see \cite{Benincasa:2022gtd}. It has been suggested that this property is intimately tied to the existence of a plethora of extreme states within the Bunch-Davies probability distribution \cite{Anninos:2011kh,Roberts:2012jw}. In order to get an intuition, we can think about the general picture of the evolution of a scalar mode in an expanding space-time. Its fluctuations can become larger than the horizon and decohere. Such {\it classicalised} modes can fluctuate and undergo the same process as before. This implies that the late-time boundary of the space-time gets populated by a plethora of configurations so that a given mode does not have a definite expectation value. Said differently, the state space at late-time is rich. This also pairs with the known fact that light states in an expanding universe can have late-time correlations that grow as the distance between two configurations grows, and hence {\it globally} cluster decomposition does not hold. What does this really mean? In order to answer this question, it is necessary to sharpen it, turning it into a question about the structure of the state space. A similar situation occurs in the context of spin glasses and it was elucidated by introducing a new order parameter, the mean square displacement distribution (or, equivalently, the overlap distribution) to have a measure of the similarity among the states showing the existence of multiple extreme states that had a hierarchical, ultrametric, organisation in a tree-like structure \cite{Parisi:1983dx, Mezard:1986mpv, Rammal:1986zz}. The presence of multiple extreme states signals the existence of {\it memory} in the system, {\it i.e.}  correlation between the root and the states as the boundary of the tree is approached. It turns out that such {\it observables}, {\it i.e.} the mean square displacement distribution and the overlap distributions, can also be defined in the expanding universe setting, producing the same type of conclusion as in spin glasses: the Bunch-Davies state for light scalars shows multiple extreme states \cite{Anninos:2011kh, Roberts:2012jw, Shaghoulian:2013qia, Benincasa:2026bcl}.

Importantly, the mean square displacement distributions offer a concrete example of {\it global} observables. Notice that our ability of defining global observables will determine our ability and depth of understanding the dynamical (and quantum) effects of gravity: the usual correlators turn out {\it not to be diffeomorphism invariant}, a basic requirement for having a consistent description of gravity.

Finally, this disordered nature turns out to be relevant as we try to shed light on the microscopic nature of gravity (in general and for an expanding space-time) as it is in the structure of the (DS)SYK model used to understand the microscopic picture of lower dimensional expanding universes. 

\begin{equation*}
    \begin{tikzpicture}
            \node[draw, rectangle, rounded corners, thick, color=blue!80, fill=blue!10, text width=14.875cm, text height=3cm] (O4) at (0,0) {};
            \node[font=\it, align=justify, text width=14cm] {%
            	\color{black}{Considering manifestly the out-of-equilibrium nature of our universe allows us to define novel observables} {and provide a new pathway in the quest for a microscopic description of expanding universes.} {So far, some lessons have been learnt from disordered systems such as spin glasses.} 
            {The chapter {\scshape Statistical systems out-of-equilibrium} \cite{Anous:2026tdu} {provides a pedagogical introduction to spin glasses, with connections to open problems in cosmology and holography.}}%
            };
    \end{tikzpicture}
\end{equation*}

\section{Information from closed and open quantum system viewpoints}\label{sec:OQS} 

Most of the recent progress in the last nine years in understanding the analytic structure of cosmological observables were focused on the so called {\it wave-function of the universe} and the correlations at the end of inflation. While the latter are the natural - with a pinch of salt -- observables as they seed the consequent evolution that generates what we observe today, the former is a more primitive object that, loosely speaking, contains all the degrees of freedom and encodes their probability distribution at the end of inflation, and whose properties are at the root of the features of any other late-time observable. In fact, they are simply related by
\begin{equation*}
	\langle f([\Phi])\rangle\:=\:\int\mathcal{D}[\Phi]\,\mathfrak{P}[\Phi]\,f([\Phi]),
	\qquad
	\mathfrak{P}[\Phi]\: :=\:\frac{\left|\Psi[\Phi]\right|^2}{\displaystyle\int\mathcal{D}\Phi\left|\Psi[\Phi]\right|^2}
\end{equation*}
where $[\Phi]$ denotes collectively {\it all the late-time configurations} at the boundary of inflation, $\Psi[\Phi]$ is the wave-function of the universe and $f([\Phi])$ is a function of such configurations. With this point of view, the evolution of the system is unitary, and the phenomenon of decoherence appears upon separation between the subsystem one would like to study and a large environment, making this subsystem look as if it were a mixed state, and the interaction with a large amount of degrees of freedom can destroy coherence in a very short time. This perspective offers the idea that while quantum effects are captured, at least conceptually, by considering everything within a closed system, classicalization emerges once one performs a separation between a system to describe and its environment from the interaction with the large amount of degrees of freedom of the environment -- {\it i.e.} by considering an open system.

Furthermore, from an observer -- and observational -- point of view, we do not perceive the universe as a closed system. Both an inflationary observer -- {\it i.e. an observer who lives in an inflationary space-time} --, and us today, are surrounded by a cosmic horizon, so we have access just to a finite amount of data, implying that any measurement carries with it a finite error. So, in order to be able to talk about the universe as a closed system, it is required either that it is assumed that at late enough times, the universe becomes infinitely large and we can access an arbitrarily large amount of data (which is not precisely the universe we live in), or that we can focus on certain degrees of freedom and in an approximation for which they are everything and we can follow them until the end of the system evolution.

There is also another point to be made. Our understanding (of physics here specifically, but actually more broadly) is really {\it effective}, {\it i.e.} we can access just some set of degrees of freedom and also interact with those emergent features from underlying degrees of freedom that we cannot resolve at the energies we are studying our system. So whenever we can have under control these emergent properties or we can work in a regime in which they can be neglected, then also the closed system perspective can hold, and if it fits our purposes, it is possible to parametrise our ignorance of such degrees of freedom by introducing new operators consistent with the symmetries of the problem and whose relevance is given in terms of Wilson coefficients -- according to the standard effective (field)\footnote{Here, the word {\it field} has been put in parenthesis as it is not really conceptually necessary. Although it is true that the most common language to describe physical phenomena at high energy is field theory, it is really {\it just} a framework, very powerful, but nevertheless a framework. At least perturbatively, it is possible to define a theory from some main basic assumptions and dealing just with physical degrees of freedom. The effective corrections can be added directly by constructing the couplings in terms of symmetries, rather than building field theory operators, as it is becoming standard practice in particle physics.} theory practise.

Hence, given our ignorance about the actual degrees of freedom in cosmology, an alternative approach is to describe cosmological problems as an open quantum system: the precise question we ask determines how the problem is divided into system and environment, for which we can then {\it effectively} parametrise our ignorance about the microscopic degrees of freedom in the environment. Systems of this type are typically {\it out of equilibrium} and can be studied via in-in correlation functions, which are of the same type as the ones whose analytic structure we have been after. 

In recent years, a great deal of progress has been made in adopting this open quantum system point of view, at first as an attempt to provide a complete framework for addressing the long standing issue of the infra-red divergent behaviour of the late-time correlation functions \cite{Burgess:2015ajz, Cespedes:2026fdp}, then to better frame the phenomena of decoherence and the possibility of recohering, getting to a first formulation of an open effective field theory of inflation \cite{Colas:2022hlq, Colas:2022kfu, Colas:2024xjy, Burgess:2024eng, Salcedo:2024smn, Colas:2024ysu, Burgess:2024heo, Salcedo:2024nex, Kranas:2025jgm, Salcedo:2025ezu, Colas:2025ind, Cespedes:2026fdp, Lausdei:2026iwe, Christie:2026dwx}. 

\begin{equation*}
    \begin{tikzpicture}
            \node[draw, rectangle, rounded corners, thick, color=blue!80, fill=blue!10, text width=14.875cm, text height=3cm] (O3) at (0,0) {};
            \node[font=\it, align=justify, text width=14.75cm] (O3txt) at (O3) {%
            \color{black}{Open quantum systems provide a more realistic -- from the point of view of an observer in an} {expanding universe -- approach to cosmology than the idealized, and conceptual, closed one, yet it is complementary to it}. {They represent one more smoking gun that a larger set of ideas from statistical systems out of equilibrium} {can help to shed light on the deepest questions in cosmology.}
            {The chapter {\scshape Open effective field theories} \cite{Colas:2026tdu} offers a pedagogical discussion of the most recent} {developments in applying these ideas to field theories, inflation and gravity}%
            };
    \end{tikzpicture}
\end{equation*} 

\section{A pathway to quantum features of cosmology} 

The questions {\it what should we compute?} and {\it which observables should we consider?} are crucial to address if we would really like to find a microscopic theory for inflation and, more generally, for our universe. And to this purpose it is important to adopt all possible points of view, from the Lorentzian, that is typical when in-in correlators and the Bunch-Davies wave-functions are considered, to the Euclidean, in which one can compute the partition functions. As we saw in Section \ref{sec:Dis}, sharpening the question and taking seriously the out-of-equilibrium nature of our universe can lead to further handles on the issue. However, such a question is not completely unrelated to the following one: {\it Which degrees of freedom should we consider?}. Posed in this way, this question is not very sharp: if we consider the whole early universe, we do not really know which is the complete set of degrees of freedom (which is the main argument for the open quantum system approach), and when we get closer to the Planck scale, we have even less of an idea of how to set up the problem. However, we do not need to get that far: a good playground is also simply the realm of quantum field theory in a fixed (expanding) background. Even in this framework things are pretty messy, but here the questions can start to be sharpened. Taking the simplest case of a de Sitter universe, the degrees of freedom are determined by the unitary irreducible representations of the de Sitter group (more generally, of the isometry group of the relevant space-time), which then get repackaged in a redundant way in fields. So, the first important step is to really get a deeper understanding of QFT in de Sitter. Even with the degrees of freedom more under control in this QFT in de Sitter approximation, there is a further problem that kicks in: the usual observables defined in this language, which we are accustomed to dealing with, live in position space (momentum space, or any other space related to them by a transform). However, whenever we decide to turn on gravity dynamically -- even if just perturbatively -- these are no longer good observables as they are not diffeomorphism invariant. In this sense, it is really suitable to develop a {\it quasi-local}  understanding of gravity in an expanding universe, modelling the actual situation that sees us measuring information coming from far away in the universe and turning it into a map of the large scale structures.

\begin{equation*}
    \begin{tikzpicture}
            \node[draw, rectangle, rounded corners, thick, color=blue!80, fill=blue!10, text width=14.875cm, text height=3cm] (O5) at (0,0) {};
            \node[font=\it, align=justify, text width=14.75cm] (O5txt) at (O5) {%
            	\color{black}{It is fundamental to have a better understanding of degrees of freedom and observables,} 
            {The chapter {\scshape Features of a $\Lambda>0$ universe \cite{Anninos:2026tdu}} {provides a pedagogical discussion of some aspects of the foundation of quantum field theory in de Sitter, digging also into features of the fixed background, going on to quantum aspects via the gravitational path integral, and the definition of quasi-local observables.}}%
            };
    \end{tikzpicture}
\end{equation*}


\section{Outlook}\label{eq:Out}

\begin{figure}[t]
	\centering
	\includegraphics[width=.5\textwidth, height=.5\textheight, keepaspectratio]{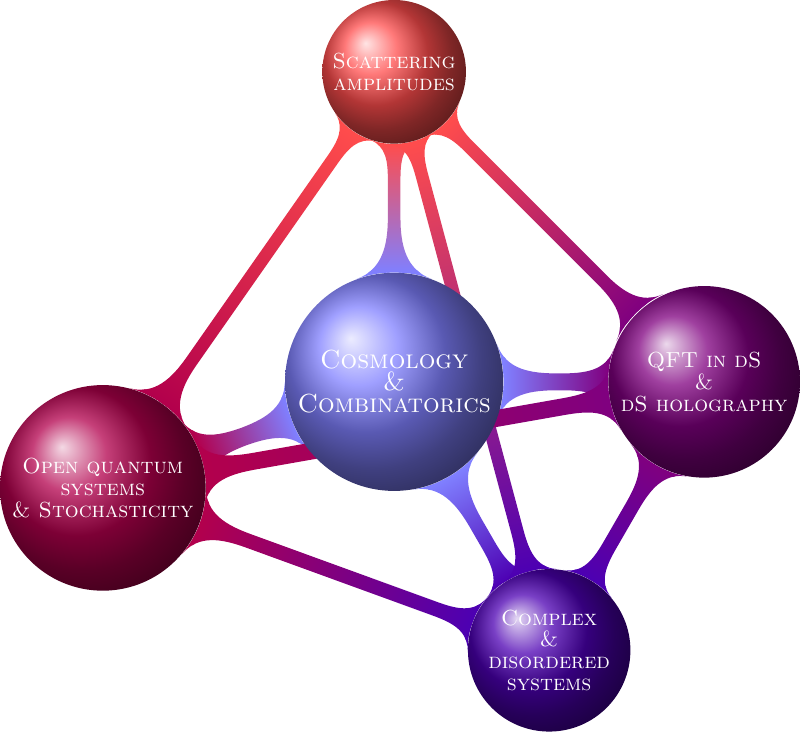}
	\caption{{\it Interplay for cosmology --}  Ideas and tools from scattering amplitudes, open quantum systems, QFT in dS and statistical systems out-of-equilibrium present a rich set that predictably has consequences in our understanding of cosmology. This graph enriches these connections based on recent progress.}
\end{figure}

The set of lecture notes collecced in \cite{Benincasa:2027tdu} provides a pedagogical introduction to four apparently disconnected topics -- open quantum systems, scattering amplitudes in particle physics, foundation of de Sitter physics and statistical systems out-of-equilibrium -- that find a common narrative in cosmology. They dealt with specific aspects of these subjects that in recent years turned out to provide a fresh look at open problems in cosmology. However, these four topics, broadly understood, merge nicely together as they can offer answers to different sets of cosmological questions, but also they can give rise to a novel language, to a novel framework to understand the physics of the early universe.

One important point that was made across the lectures is that we lack a complete knowledge of what the degrees of freedom in play are, which ranges from partial ignorance when we treat the inflationary scenario, to complete ignorance when we want to face the physics at the Planck scale. However, the situation is actually fuzzy even in the more comfortable context of quantum field theory in de Sitter. In this case, the degrees of freedom are given by the unitary irreducible representations of the isometry group, which then get repackaged in a redundant way in terms of fields. This repackaging is redundant because fields are not defined uniquely, a feature that can manifest itself in the fact that they can contain a larger set of degrees of freedom that are then eliminated via gauge transformations -- said differently, the request of gauge invariance of an observable comes from the fact that it is necessary to get rid of unphysical degrees of freedom. Let's clarify this point by taking an example of what happens in particle physics and in quantum field theory in flat space-times. 

As we just outlined in de Sitter case -- and it is indeed the same for any space-times -- the degrees of freedom we can have are provided by the unitary irreducible representation of the isometry group, in the flat-space case the Poincar{\'e} group, and the fields are, let's stress it again, a redundant way of repackaging them. This field-repackaging step can be avoided if we work directly at the level of {\it observables}. But what are they? One could start by just asking the question about which Poincar{\'e} invariant object can be defined including two, three and more of these unitary irreducible representations, which is simultaneously covariant under the little group -- this last feature is what distinguishes the unitary irreducible representations from one another and the degrees of freedom in the process. In the case of the Poincar{\'e} group, these representations are finite dimensional\footnote{There are actually two further irreducible representations that are infinite dimensional, but have been historically discarded on the basis of heuristic arguments: one is actually a vacuum state, which is labelled by the Casimir operators of the Lorentz subgroup, and the other gives rise to the so called continuous spin particles. While the former could acquire some relevance in a better understanding of the infrared behaviour and the definition of an infra-red safe S-matrix, the latter got a revived interest as legitimate states recently \cite{Schuster:2023xqa, Schuster:2024wjc, Bellazzini:2024dco, Kundu:2025fsd, Kundu:2025mzm, Basile:2026ntj}} and these objects we are looking for can be constructed out of polynomial of spinors \cite{Gelfand:1966ggv}, with the spinors parametrising the kinematic space. This perspective allowed us to fix the simplest physical processes involving three particles \cite{Benincasa:2007xk, Arkani-Hamed:2017jhn}. They can be used to classify fundamental theories and as building blocks to construct processes involving a larger number of particles, supplementing the symmetry information with the notions of cluster decomposition and unitarity, that allows to glue the three-particle amplitude into higher-point amplitudes. This provides a way to define perturbation theory in particle physics without the notions of fields nor Lagrangians, and provides natural consistency conditions on the physics \cite{Benincasa:2007xk, Arkani-Hamed:2017jhn}. 

It would be ideal to reach a similar understanding in de Sitter. For this reason we thought it was fundamental to go through the irreducible representations of the de Sitter group and we foresee novel developments on the lines we just discussed in the near future. One important point is that this strategy allows us to naturally define one (or more?) classes of observables. Which brings us back to the important question {\it What should we compute?}.

It is important to mention that there have been attempts to define a concept of S-matrix in de Sitter space-time \cite{Marolf:2012kh, Melville:2023kgd, Melville:2024ove, Goyal:2025hdm}, which try to define an object that can share similar properties of the flat-space S-matrix and/or make more transparent how to implement those ideas that worked so beautifully for the flat-space S-matrix. 

Another flat-space observation is that several different observables can be related by analytic continuation \cite{Caron-Huot:2023vxl, Caron-Huot:2023ikn, Caron-Huot:2025ymc}. Understanding and analysing the analytic continuation of the known late-time observables in de Sitter/cosmology, can trace a route to contribute to address this question. 

In the quest for addressing this question, it is also important to have in mind the existence of infra-red divergences for light states, which are due to the long-wave modes getting piled up at future infinity by the expansion of the universe, and which could in principle signal a pathology of a theory in de Sitter or even an instability of de Sitter space itself \cite{Ford:1984hs, Antoniadis:1985pj, Tsamis:1994ca, Tsamis:1996qm, Tsamis:1997za, Polyakov:2007mm, Polyakov:2009nq, Senatore:2009cf, Giddings:2010nc, Giddings:2010ui, Burgess:2010dd, Marolf:2010zp, Rajaraman:2010xd, Marolf:2010nz, Krotov:2010ma, Giddings:2011zd, Marolf:2011sh, Giddings:2011ze, Senatore:2012nq, Polyakov:2012uc, Beneke:2012kn, Akhmedov:2013vka, Anninos:2014lwa, Akhmedov:2017ooy, Hu:2018nxy, Akhmedov:2019cfd, Gorbenko:2019rza, Mirbabayi:2019qtx, Cohen:2020php, Mirbabayi:2020vyt, Cohen:2021fzf}. So, there is nowadays a common lore that, in a healthy theory, these infrared divergences can be re-summed and the re-summation of the leading divergences is compatible with stochastic inflation \cite{Rajaraman:2010xd, Beneke:2012kn, Burgess:2015ajz, Gorbenko:2019rza, Baumgart:2019clc, Cespedes:2023aal}. However, despite the proposals in the realm of soft de Sitter effective field theory \cite{Cohen:2020php, Cohen:2021fzf, Beneke:2026rtf} and open effective field theory \cite{Burgess:2015ajz, Cespedes:2026fdp}, there is no complete framework yet for a full-fledged incorporation of this IR divergences and, especially, there is no definite answer to the questions {\it Can we define from first principle IR-finite correlators?} {\it Is it possible to define IR-finite observables?}. The two effective field theories just mentioned suggest that the answer to the first question is indeed positive, as a first work on consistent subtractions for arbitrary graphs \cite{Benincasa:2024rfw}\footnote{In \cite{Benincasa:2024rfw}, it is argued that it is possible to define a perturbative order-by-order cancellation for graphs appearing in the Bunch-Davies wave-function and correlators expansion by observing the IR structure that different graphs share, but that their IR-finite combination would define the perturbative expansion for a different observable, along the same lines of what happens in flat space where the IR-finite combination of Feynman graphs \cite{Anastasiou:2018rib} are related to a Wilson loop with a local operator insertion \cite{Alday:2011ga, Engelund:2011fg, Alday:2012hy, Engelund:2012re, Arkani-Hamed:2021iya}. Another subtraction procedure was proposed in \cite{Bzowski:2023nef} in the realm of holographic renormalisation, assuming that there exists a weak-strong duality between the gravitational theory in de Sitter and a CFT on the de Sitter boundary and hence that the IR divergences of the former can be treated as UV divergences of the latter.} seem to suggest that also the answer to the second question is positive.

Furthermore, for the quest to be considered completed, we would need to understand how to perform the step from these observables that can make sense as long as the approximation of a fixed background and perturbation around it holds, to observables which are diffeomorphisms invariant, that are crucial when gravity is dynamical -- in these lectures we learnt about an intermediate step, {\it i.e.} the definition of {\it quasi-local} observables.

Finally, in this context there are two main lessons to draw also from the spin-glass discussion in these lectures. The first one is that depending on the question we would like to address, it is necessary to first ask {\it What should we compute?} -- this is something that we need to get accustomed to asking, rather than just {\it shut up and compute}. Despite motivated by observations and by the evolution of our universe, the in-in correlation functions and the wave-function/reduced density matrix do not allow us to provide a complete microscopic picture of our universe. Information on the state-space structure and the memory in the system, could be extracted by introducing new ideas. The second one is that taking more seriously that our universe is an out-of-equilibrium system can be useful to address this question and extract physics that otherwise would be more difficult to learn.  

Hence, we can see how the interplay among the apparently different topics of scattering amplitudes, open quantum systems, QFT in de Sitter/de Sitter holography and complex systems, can have still more to say, and it is a direction which is worth pushing and whose advances are worth teaching. In fact, a second edition of the school whose lectures this book is based on, -- again named \href{https://indico.global/event/16455/}{\it The Disordered Universe 2026} -- already took place in the same venue, the {Espazo La Ricotta} in As Barreiras (Castro Caldelas, Spain), dealing with different aspects of these topics and their interplay.

Let us conclude by stressing how this is a long term direction. As we all well know, QFT and GR are the most common language used to study and understand particle physics and cosmology, and an effective description of particle physics without and with gravity respectively. The latter was proven to be 2-loop non-renormalisable \cite{Goroff:1985sz}, and 1-loop non-renormalisable in presence of matter \cite{tHooft:1974toh}, implying that its predictive power decreases as we go higher in perturbation theory until it becomes totally unpredictive. Said differently, the framework of GR does not allow us to extrapolate our knowledge of gravity from low energies to arbitrarily high energies. Furthermore, as the Planck mass scale is approached, no local observables can be defined, implying the breakdown of (at least) one of our fundamental pillars. 

Furthermore, the very same principles that seem fundamental at accessible high energies, become approximate at cosmological scales: the expansion of the universe, both during inflation and the current one, break Lorentz invariance; the natural global observables do not have a natural notion of unitarity as they live on space-like surfaces with the time evolution being integrated out, and furthermore even in the approximation of a QFT in an expanding background, the global observables with sufficiently light states do not satisfy cluster decomposition, contrarily to what happens in flat-space. Hence, in order to have a deeper understanding of fundamental interactions in and beyond the accessible high energy regime, it becomes crucial to understand the principles ruling the physics in an expanding background and how they are linked to the flat-space ones. And, as we outlined earlier, understanding processes in expanding backgrounds can shed light on the physics at such {\it ultra high energies}. It is really fascinating that unveiling the mysteries of the physics in an expanding universe can have consequences at (almost) all energy scales.

All this points to the necessity of developing a new language for describing ultra high energy phenomena, such that the principles that seem fundamental at lower scales, naturally emerge from the principles of such a novel formulation. In this direction there are two concrete proposals: one that pursues ``Parke-Taylor''-like simple formulas in de Sitter, and the other one that aims to define novel mathematical objects from mathematical first principles from which both the physical principles and the physical observables emerge. 

The former recently found its codification in the {\it cosmological Grassmannian} \cite{Arundine:2026fbr, Arundine:2026myr}, which represents Bunch-Davies wave-function coefficients in an auxiliary space, the orthogonal Grassmannian, where simpler structures emerge. However, so far it has only allowed us to produce tree-level formulas for three and four external states, and for gravitons it needs more systematisation. Again, it is the beginning of a new and exciting road. 

The latter description is instead in terms of {\it cosmological polytopes} \cite{Arkani-Hamed:2017fdk, Arkani-Hamed:2018bjr, Benincasa:2018ssx, Benincasa:2019vqr, Benincasa:2020uph, Benincasa:2020aoj, Benincasa:2021qcb, Benincasa:2022gtd, Benincasa:2022omn, Benincasa:2024leu}, which are combinatorial objects in $1-1$ correspondence with cosmological Feynman graphs associated to scalars with polynomial interactions. This also allows a systematic study of the asymptotic behaviour of the cosmological processes, including their IR behaviour in full generality \cite{Benincasa:2024lxe} and identifying cancellation patterns among graphs \cite{Benincasa:2024rfw} -- this provides a new handle for the formulation of a general and consistent framework for the IR effects and for the definition of IR finite quantities. 

The $1-1$ correspondence between cosmological polytopes and graphs is at the same time a blessing and a curse. It is a blessing because it allows one to make general statements not only about the classes of theories the graphs directly describe, but more generally about those processes that can be expanded in such graphs. It is a curse because, when considering processes involving spin-1 states or, more importantly, gravitons, then it does not make it easy to get rid of the gauge redundancies. However, the existence of such description, whose definition -- it is important to stress -- is intrinsic, provides a language that naturally unifies the physics in flat and expanding space-times, and constitutes a handle on understanding their relation, as well as on novel computational techniques for both sets of processes. Very importantly, precisely because now physical processes can be described in terms of cosmological polytopes, we have at our disposal novel rules and novel operations we can do on such structures and from which we can extract new physical results. Needless to say, because of the curse mentioned earlier, it is clear this combinatorial description has just scratched the surface. However, it makes clear that a language different from field theory is not only needed to achieve a deeper understanding of fundamental physics, but is also possible. Furthermore, such a framework returns the relevant observables -- both in expanding and flat space-times -- directly in terms of boundary data providing a handle in the quest for a microscopic description via a novel viewpoint on the realisation of holography. 

These lecture notes, and more generally the {\it Disordered Universe} program, opened a novel long term direction which promises to strengthen and deepen the links among the four fields it incorporates, producing new ideas and new results to be disseminated in new editions of the Summer School on which this book is based. 

\begin{acknowledgements}
	-- This article and the lecture notes associated to it have been made possible by the great enthusiasm and effort of all the participants and lecturers, who contributed beautifully to it, bringing forth a completely novel product, and not just for the novelty in and of itself of its content: Alessandro Fumagalli, Alessandro Palazio, Andrew McLeod, Chandra Chowdhuri, Chantelle Esper, Ciaran McCulloch, Daniel McLoughlin, Dionysios Anninos, Francesco Paronetto, Jonah Stalknecht, Leonardo Stanzani, Lucas Vicente Garc{\'i}a-Consuegra, Mart{\'i} Berenguer Mim{\'o}, Mattia Capuano, Mattia Varrone, Omkar Shetye, Polina Petriakova, Rafa Carrasco Carmona, Rajat Sharma, Sam Brian, Silvia Pl{\'a} Garc{\'i}a, Tarek Anous, Thomas Colas, Tommaso Marini, Weam Abou Hamdan, Yakov Landau, Yong Sheng Yap.

However, there have been a lot of invisible -- from outside -- hands that have made the school possible, and hence this paper and the lecture notes. Those are the hands of friends who crucially contribute to the set up of the venue, but also to keep it running when Murphy's law hit hard. Those are the hands of: Alex, Alfonso, Ana, Anna, Bel{\'e}n, Borja, Carolina, Dani, Diana, Gloria, Jorge, Jos{\'e} (Joey), Lois, Lotus, Paula, Saxo, Steve, Subodh, Vero. A special thanks goes to Anna, Borja and Paula, the team that took care of pretty much everything during the two weeks of the school, and Alex and Dani who constituted a sort of rescue team for whatever was needed. Behind the realisation of a successful idea, there is never a single person or a very reduced number of people, rather a series of conditions, that take the form of those invisible hands but also of unexpected support. Indeed, in this category falls the support of all the neighbours in Poboeiros, and everyone else that welcomed with unexpected enthusiasm both the summer school in and of itself and all the open events that it carried along.

Indeed a big thanks goes to the {\it Asociaci{\'o}n Cultural ``La Ricotta''}, in the persons of Cris, Ernesto and Kepa.

It is also a pleasure to thank Javier Mas for unconditional support in giving form to the idea behind the school, and not only that, and for being able to drag the IGFAE into it, also thanks to the enthusiasm of Ricardo Rodr{\'i}guez Fern{\'a}ndez and Manuel Pan Rey. We also thank the event staff of Scuola Normale Superiore, particularly Elisa Guidi, for help in logistics.

P.B. would also like to warmly thank Dionysios Anninos, Tarek Anous and Jorge Kurchan for discussions and setting all together the stage for the general scientific idea behind the {\it Disordered Universe} program, that evolved into the current status and found its first incarnation in the \href{https://indico.sns.it/event/91/}{{\it La Ricotta} Summer School: {\it The Disordered Universe 2025}}. 

The \href{https://indico.sns.it/event/91/}{{\it La Ricotta} Summer School  {\it The Disordered Universe 2025}} received financial support from the ERC grant NOTIMEFORCOSMO (agreement n${}^{\circ}$ 1126304)), by Scuola Normale and by the Instituto Galego de F{\'i}sica das Altas Enerx{\'i}as (IGFAE-USC/Xunta de Galicia) through the CIGUS network and EU cofinancing program.

\end{acknowledgements}

\appendix

\section{The Disordered Universe Summer School(s) and more}\label{app:DU}

It is an interesting time to work in fundamental physics. Both particle physics and cosmology have their “standard models” which pass most stress tests with flying colours. Having no clear path forward from experiment or observations, theorists must explore many directions. The reason is twofold. First, theorists aim to produce more exquisite and accurate predictions from our current models of the universe, to suggest new observables to look for new physics, or to unveil the microscopic ingredients that make up our universe. Second, this exploration often opens new bridges with other parts of physics, unifying apparently disconnected areas, offering new perspectives, and enabling progress. Our goal with this school was to bring together experts in topics that are close, but not so close to each other, where we feel that amalgamation will soon happen—hopefully by the hands of the young participants.

The \href{https://indico.sns.it/event/91/}{{\it La Ricotta} Summer School {\it The Disordered Universe}} was focused on a few directions that look promising when it comes to understanding the origins and fate of our universe, at its largest scales:
\begin{itemize}
	\item[\ding{111}] Understanding the methods of quantum field theory in setups with dissipation and particle production is of clear importance to cosmology;
	\item[\ding{111}] In order to better compute cosmological correlators, much expertise from particle physics is needed and can be imported, including the profound consequences of unitarity and locality in scattering amplitudes;
	\item[\ding{111}] Disordered systems like spin glasses have enormous entropy and frustration, providing interesting theoretical laboratories that could mimic some features of expanding space-times, which have a large horizon entropy; 
	\item[\ding{111}] The simplest toy model of the expanding universe -- de Sitter space -- still harbors many puzzles related to its proper treatment in quantum field theory, and in particular in quantum gravity.
\end{itemize}

The lecture notes associated to the present introductory paper is not meant to be just one more collection academic lecture notes. Rather it is at the same time the end point of a self-contained pedagogical path, and the starting point for a long term scientific and training vision.

It is the result of the collective effort of all school participants, lecturers and organisers: each chapter -- other than the introduction and the conclusion, which have been taken care of by the school organisers -- reflects the content of one of the courses of the school and is the outcome of the collaboration among a spontaneously-formed subgroup of the participants, supervised by the relevant lecturer, and everyone guided and coordinated by the organisers, who acted as editors of this book. This pedagogical path starts with the design of the school itself, which brings together the frontier developments in four apparently distinct research topics -- open quantum systems, scattering amplitudes in particle physics, quantum field theory in de Sitter, and statistical systems out of equilibrium --, to address longstanding open questions in cosmology, taught by leading scientists in their respective fields. Each course had overall seven-and-a-half hours of lectures, complemented by five hours of tutorials, during which the participants had to solve in groups problems posed by the lecturers and, then, share their work and solutions with the rest of the participants. The core idea is that the participants had to be involved actively in the school, and also have the opportunity to fix in their minds the main ideas of each course while collaborating. Sharing and collaborating has been the {\it leit motiv} of the school: collaborating and sharing science in the lecture room as part of the formal courses; collaborating and sharing science informally in the free time, from meals to the leisure time spent in the venue located in the countryside of the {\it Ribeira Sacra} region, and under its precious night sky, which came with the camping-tent accommodation on offer; collaborating and sharing science with the lecturers, made easier by the informal setting that facilitated spending time together; collaborating and sharing with people from the surroundings, invited at the venue when its doors got opened to host public events in which the participants themselves were involved. 

The lecture notes hence delivers not only the scientific content of the courses that characterised the school and thus the novelty of the research line given by their intersection, but also -- and especially -- {\it how} it got generated, the informal, inclusive and collaborative approach behind the school. In this sense it is the end point of a self-contained pedagogical path: it has been the last chance to stop, rewind, process individually and share among all people involved in the school those two weeks of scientific (and human) experience, and sharing it further with the rest of the academic community. These lecture notes delivers the scientific content of the school and its methodology.
 
However, it is also the starting point for a long-term scientific and training vision, {\it i.e.} optimise the resources we can have for the organisation of scientific events in order to
\begin{itemize}
	\item[\ding{111}] reduce (with the aim to suppress) economical barriers to the participation of graduate students, by offering as much as possible;
	\item[\ding{111}] create a scientifically stimulating environment which can be pedagogically effective, setting up time for both individual and collective rethinking of the topics of the lectures;
	\item[\ding{111}] provide a working example of collaboration over competition, which is made concrete by the tutorials, the lecture notes and the book \cite{Benincasa:2027tdu};
	\item[\ding{111}] have a backreaction on the surroundings where the scientific event takes place, via a series of public activities that take place during the event itself but also during the rest of the year -- the non-profit cultural association {\it La Ricotta} which takes care of the venue and contribute to design the scientific plan of the events, pursues the diffusion of the {\it scientific habit of mind} in rural areas, especially where the venue is located.
\end{itemize}

This is a model that points towards economical and environmental sustainability, scientific excellence, pedagogical effectiveness, and short- and long-term social impact, while simultaneously being accessible to any graduate student and early career postdoc, provided they fit into the scientific programme. Importantly, in order for all this to happen, it should keep its current size but can be easily replicated by others, in the interest of resources, scientific pedagogy and society. In this sense, these lecture notes represent the beginning of a journey that starts and continues in As Barreiras, but hopes to have an influence on how scientific events are conceived.

\bibliographystyle{utphys}
\bibliography{2025-DU-Intro-Biblio}	

\end{document}